\documentclass[a4paper, 11pt]{article}

\usepackage[T1]{fontenc}

\usepackage{amssymb}        
\usepackage{latexsym}
\usepackage{times}
\usepackage{mathptm}

\date{}
\usepackage{graphicx}

\newcommand{\eps}{\epsilon}

\newtheorem{fact}{Fact}

\newtheorem{lemma}{Lemma}
\newtheorem{theorem}{Theorem}
\newtheorem{corollary}{Corollary}
\newtheorem{definition}{Definition}
\newtheorem{invariant}{Invariant}

\begin{document}
\title{A  $7/9$ - Approximation Algorithm for the Maximum Traveling Salesman Problem}

\author{%
Katarzyna Paluch\thanks{Institute of Computer Science, Wroc{\l}aw University. Work partially done while the author
was in Max Planck Institute for Computer Science, Saarbruecken}  
\and Marcin Mucha\thanks{Warsaw University} 
\and Aleksander Madry \thanks{MIT}
}

\maketitle{}      
\thispagestyle{empty}

\section{Introduction}
The traveling salesman problem is one of the most famous and heavily researched problems in computer science.
The version we deal with in this paper is the Symmetric Maximum Traveling Salesman Problem, which is defined
as follows.
For a given complete undirected graph $G$ with nonnegative weights on its edges, we wish
to find a tour of the graph of maximum weight. The tour of the graph is a simple cycle that contains each vertex from
$G$. 
In 1979 Fisher, Nemhauser and Wolsey \cite{Fish} showed that the {\em greedy}, the {\em best neighbour} and the {\em $2$-interchange} algorithms have approximation ratio $1/2$.
In \cite{Fish} the {\em $2$-matching} algorithm  is also given, which has a guarantee of $\frac{2}{3}$. In 1994 Kosaraju, Park and Stein \cite{Kos} presented an improved algorithm
having ratio $\frac{19}{27}$ \cite{BH}. 
In the meantime in 1984  Serdyukov \cite{Ser} presented (in Russian) a simple (to understand) and elegant $\frac{3}{4}$-approximation algorithm. The algorithm
is deterministic and runs in $O(n^3)$. 
Afterwards, Hassin, Rubinstein (\cite{HR}) gave a randomized algorithm having {\it expected} approximation ratio at least $\frac{25(1-\eps)}{33-32\eps}$  and running in $O(n^2(n+2^{1/\eps}))$, where $\eps$ is an arbitrarly small constant.
The first deterministic approximation algorithm with the ratio
better than $\frac{3}{4}$ was given in 2005 by Chen, Okamoto, Wang (\cite{Chen}), which is a $\frac{61}{81}$ approximation and a nontrivial derandomization of the algorithm from \cite{HR}. It runs in $O(n^3)$.

{\bf Related work} For the asymmetric version of Max TSP, the best approximation is by Kaplan, Lewenstein, Shafrir, Sviridenko (\cite{Svir}) and has ratio $\frac{2}{3}$. If aditionally in graph $G$ triangle inequality holds, we get two (symmetric and asymmetric) metric
versions of the problem. The best approximation bounds for them are $\frac{7}{8}$ (\cite{HRm}) and $\frac{10}{13}$ (\cite{Svir}), both of which have been  improved  by Chen and Nagoya  in \cite{Chenn}. The latest improvements are by Kowalik and Mucha (\cite{KM} and \cite{KM2}) and equal, respectively for an asymmetric version $\frac{35}{44}$ and for the symmetric version $\frac{7}{8}$. All four versions of Max TSP are MAX SNP-hard (\cite{Eng},\cite{EK},\cite{PY}). 
A good survey of the maximum TSP is \cite{Bar}.

{\bf Our results}
We give a fast deterministic combinatorial algorithm  for the Symmetric Maximum Traveling Salesman problem, with the approximation guarantee equal to $\frac{7}{9}$. 
To achieve this, we compute the graph described in the following theorem, which is proved in Section \ref{upper}.
\begin{theorem} \label{main} Given a complete graph $G$ with nonnegative weights on the edges,
we can compute a multisubgraph $H=(V,E_H)$ of $G=(V,E)$ such that
$H$ is loopless, $4$-regular, each $e \in E_H$ has the same weight as in $G$,
there are at most two edges between a pair of vertices, each connected component has at least $5$ vertices and its weight is at least $\frac{35}{18} opt.$ ($opt$ denotes the weight of an optimal tour.)  
\end{theorem}

The method used in this theorem  is new and can be used for any optimization problem for which a cycle cover of minimal/maximal weight is a lower/upper bound on the optimal value of the solution. (In Appendix the use of this method
for $(1,2)$-TSP is attached.)
In the proof we exploit the fact that the tour of the graph is a cycle cover of $G$ or in other words a simple perfect $2$-matching. Thus a maximum weight cycle cover $C$ of $G$ is an upper bound on $opt$. The tour of the graph in turn is a somewhat special cycle cover, it has some properties we can make use of and the
the notion from the matching theory that turns out to be particularly useful is that of an alternating cycle.

Next in the proof of Theorem \ref{koloro} we show how to extract from $H$ a tour of weight at least $\frac{2}{5} \cdot \frac{35}{18} opt$.
 \begin{theorem} \label{koloro}
If we have a loopless $4$-regular graph $H=(V,E_H)$ with nonnegative weights on the edges that can contain at most two edges
between a pair of vertices  and such that its every connected component has at least $5$ vertices,
then we can find such a subset $E'$ of its edges that $w(E') \leq 1/5 w(H)$ and such that we can $2$-path-color the graph $H'=(V,E_H \setminus E')$.
\end{theorem}  
To $2$-path-color the graph means to color its edges into two colors so that no monochromatic cycle arises.  
The outline of the proof of this theorem is given in Section \ref{heavy}.
The whole algorithm  runs in time $O(n^3)$, where $n$ denotes the number of vertices in $G$. 
The estimation of the approximation ratio is tight. The obstacle to $4/5$-approximation is that we are not able to construct an exact gadget for a square. Gadgets for squares are described in Section \ref{upper}.

For comparison, let us note, that in the case of the Asymmetric Max TSP, which is considered
in \cite{Svir}, the authors compute a $2$-regular loopless graph $G_1$ (which is a multisubgraph of $G$), whose all connected components contain at least $3$ vertices and such that its weight is at least $2 opt$. Next a tour of weight at least $\frac{1}{3} 2 opt$ is extracted from $G_1$. However obtaining graph $G_1$ in \cite{Svir} 
is not combinatorial. It involves using a linear program that is a relaxation of the problem of finding
a maximum cycle cover which does not contain $2$-cycles. Next  scaling up the fractional solution by an appropriate integer $D$ (which is a polynomial in $n$) to an integral
one, which defines a $d$-regular multigraph, from which  a desired graph $G_1$ is obtained. The running time needed
to compute $G_1$ is $O(n^2D)$.

\section{Upper bound} \label{upper}
Let $G=(V,E)$ be a complete graph with nonnegative weights on the edges, in which we wish to find a traveling salesman tour (a cycle containing all vertices from $V$) of maximum weight. 
Let $T_{max}$ denote any such tour and $t_{max}$ its weight.
  
The weight of the edge $e=(u,v)$ between vertices $u$ and $v$ is denoted by $w(e)$ or $w(u,v)$.
By $w(E')$  we denote the weight of the (multi)set of edges $E' \subset E$, which is defined as $\sum_{e \in E'} w(e)$. The weight of the graph $G$ is denoted as $w(G)=w(E)$. 

One of the natural upper bounds for $t_{max}$ is the weight of a maximum weight cycle cover $C$ of $G$ ($C$ is a cycle cover of $G$ if each vertex of $V$ belongs to exactly one cycle from $C$). 
If $C$ contained only cycles of length $5$ or more, then by deleting the lightest edge from each cycle and patching them arbitrarily into a tour  we would get a solution of weight at least $\frac{4}{5}\ t_{max}$. $C$ however can of course contain triangles and quadrilaterals.
From now on, let  $C$  denote a cycle cover of maximum weight and assume that it contains more than one cycle. Further on, we will define the notions of a good cycle cover and alternating weight. They will be strictly connected
with $C$.

We can notice that $T_{max}$ {\it does not} contain an edge (one or more)  from each cycle from $C$.
Since, we aim at a $\frac{7}{9}$-approximation, we will restrict ourselves to bad cycles from $C$, which are defined as follows. Cycle $c$ of $C$ is said to be {\bf bad} if each edge of $c$ has weight greater than $\frac{2}{9} w(c)$.
Let us notice that if a cycle $c$ is bad, then it is a triangle or a quadrilateral. For convenience, let us further on
call all quadrilaterals {\bf squares}.
We will call a cycle cover $C'$ {\bf good} if for each bad cycle $c$ of $C$,  $C'$ does {\it not} contain at least one edge from $c$ and if it does not contain a cycle
whose vertices all belong to some bad cycle $c$ of $C$ (which means, informally speaking, that $C'$ does not contain cycles that are ''subcycles'' of the bad cycles from $C$). Since $T_{max}$ is just one cycle, it is of course good 
and the weight of a good cycle cover of maximum weight is another upper bound on $t_{max}$.
See Figure \ref{kwadtroj} for an example of a good cycle cover.

\begin{figure} 
\centering{\includegraphics{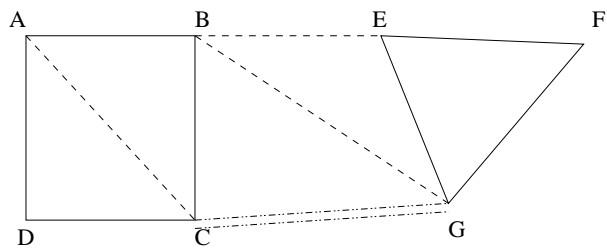}}
\caption{\scriptsize The weight of the edges drawn with a solid line is $7$, with a dashed line $6$ and the weight of edge $CG$ is $3$. The remaining (not drawn) edges have weight $0$. The cycle cover $C$ of maximum weight consists of cycles $ABCD$ and $EFG$.
The cycle cover $C_1$ consisting of cycles $ACD$ and $BEFG$ is not a good cycle cover as $ACD$ is a subcycle of $ABCD$. A good cycle cover $C'$ consists of cycle $ABEFGCD$. } \label{example}
\end{figure}

However we do not know how to find a good cycle cover of maximum weight and instead will find something
that approximates it.

\subsection{Approximating a good cycle cover}
We will construct graph $G'$ and define a special $b$-matching $B$ for it, so that $B$ of maximum weight will
in a way approximate a good cycle cover of maximum weight in $G$. 
(A $b$-matching is such a generalization of a matching in which every vertex $v$ is required to be matched with $b(v)$ edges.)   
  
Let $C'$ denote a good cycle cover of maximum weight.  
$C$ and $C'$ are cycle covers or, in other words, simple $2$-matchings ($2$-matchings and their generalizations are desribed, among others,
in \cite{Sch}).
Let us look closer at $C \oplus C'$ (i.e. the symmetric difference between sets of edges $C$ and $C'$) and get advantage from the matching theory in order to notice useful properties of a good cycle cover.

% and thus the task of finding $C'$ is equivalent to finding a set of alternating cycles with respect to $C$, such that 
%they contain at least one edge from each bad cycle from $C$ and such that the weight of $C$ is reduced as a result in the least possible way.
%Let us explain it more precisely.
First, recall a few notions from matching theory.
A path $P$ is {\bf alternating} with respect to a cycle cover $C_1$ if its edges are alternatingly from $C_1$ and from $E \setminus C_1$.
If an alternating path ends and begins with the same vertex, then it is called an {\bf alternating cycle}.
For any two cycle covers $C_1$ and $C_2$, $C_1 \oplus C_2$ can be expressed as a set of alternating cycles (with respect to $C_1$ or $C_2$).

Since $C'=C \oplus (C \oplus C')$,
$$w(C')=w(C)-w(C\cap(C\oplus C'))+w(C'\cap(C\oplus C')).$$
For convenience, we will also use the notion of {\bf alternating weight} $w'$ and define it for a subset $S$ as $w'(S)=w(S\setminus C) - w(C\cap S)$.
Using it we can rephrase the above statement as 
$$w(C')=w(C)+w'(C\oplus C').  \label{formula} \ref{formula}$$

For example in Figure \ref{example}  $(C \oplus C_1)$ is an alternating cycle $(BE,EG,GB,BC,AC,AB)$, whose alternating weight amounts to $-3$.  $(C \oplus C')$ is an alternating cycle $(BE,EG,GC,CB)$ whose alternating weight 
amounts to $-5$.

If we have an alternating cycle $A$ with respect to $C_1$, then by {\bf applying} $A$ to $C_1$ we will mean the operation, whose result is $C_1 \oplus A$. 

In view of \ref{formula} we can look at the task of finding a good cycle cover as at the task of finding a collection $A'$ of alternating cycles with respect to $C$, such that each bad cycle from $C$ is ''touched'' (i.e.some edge from a bad cycle
$c$ belongs to some alternating cycle from $A'$)
by some alternating cycle from $A'$ and the weight of $C$ diminishes in the least possible way as a result of applying $A'$ to $C$.

In the following fact we describe good cycle covers from the point of view of alternating cycles (with respect to $C$).

\begin{fact} \label{przek}
If $C'$ is a good cycle cover,
then if we decompose $C\oplus C'$ into alternating cycles, then for each bad cycle $c$ from $C$, there exists an alternating cycle  
$K_c$ containing a subpath $(v_0,v_1,v_2,\ldots,v_k,v_{k+1})$ ($k\in \{2,4\}$, i.e. a subpath has length $3$ or $5$) such that vertices $v_1,v_2,\ldots,v_k$ are on $c$, vertices $v_0$ and $v_{k+1}$ are not on $c$
and $v_1 \neq v_{k}$.  
\end{fact}

This fact follows from the definition of a good cycle cover that states that a good cycle cover does not contain cycles
that are ''subcycles'' of cycles from $C$.

Notice that in Figure \ref{example} $C_1$ is not a good cycle cover and the alternating cycle $C \oplus C_1$ contains a subpath $BC,CA,AB$ for a square $ABCD$ and it is not such as we desire as at "enters" and "leaves"
$ABCD$ with the same vertex $B$.

We define a graph $G'$ and  function $b$ for a $b$-matching in it as follows.

\begin{figure} 
\centering{\includegraphics[scale=0.7]{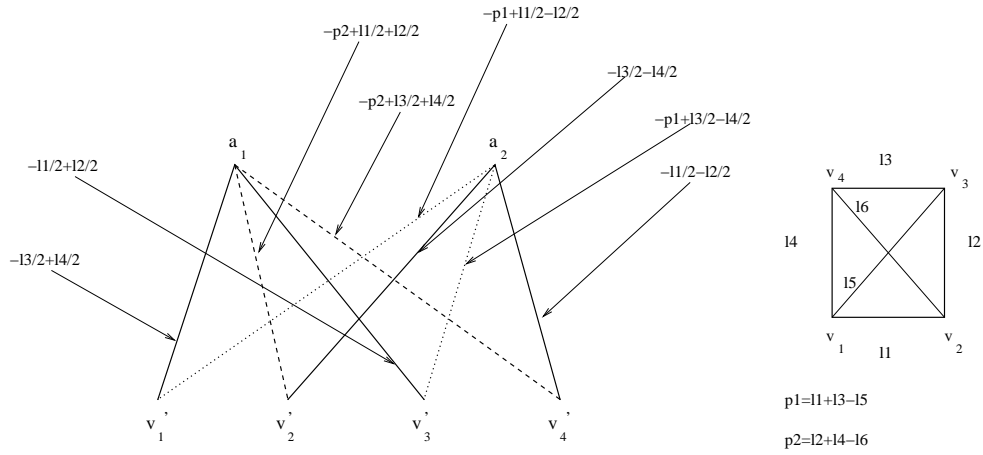}}
\caption{\scriptsize  For a bad cycle $c_i=(v_1, v_2,v_2,v_4)$, $T_i$ contains two additional verices $a_{c_1}, a_{c_2}.$ We set $b(a_{c_1})=b(a_{c_2})=1$. The fragment connected with $v_1$ and $v_2$ is the edge $(v_1,v_2)$.
There are two fragments connected with $v_1$ and $v_3$: $(v_1,v_2,v_4,v_3)$ and $(v_1,v_4,v_2,v_3)$.  
} \label{bmatch1}
\end{figure}

\begin{definition}
The construction of $G'=(V',E')$:
\begin{itemize}
\item graph $G$ is a subgraph of $G'$,
\item $V'$ consists of $V$ and also a set $S_i$ of additional vertices for each bad cycle $c_i$ from $C$: 
 $S_i$ contains a copy $v'$ for each vertex $v$ from $c_i$ and also a set of special vertices $T_i$. 
 The subgraph of $G'$ induced by $S_i$ is called a gadget $U_i$ corresponding to $c_i$.
 If $c_i=(v_1,v_2,v_3)$ is a triangle, then $T_i$ consists of one vertex $a_{c_i}$. The weight of the edge between $v_1'$  and $a_{c_i}$ is equal to $-w(v_2v_3)$ and analogously for vertices $v_2',v_3'$.
 We set $b(a_{c_i})=1$.
 The description of the gadget for a square  is given in Figure \ref{bmatch1}.
\item if $v_1$ is a vertex on some bad cycle $c$ of $C$, then $G'$ contains edges $(v_1',v_2), (v_1',v_2')$ iff $v_2$
is not a vertex of the bad cycle $c$ containing $v_1$.  The weight of these edges is the same and equals $w(v_1,v_2)$.
\end{itemize}

A $b$-matching for $G'$ is such that for $v \in V$, we put $b(v)=2$ and for $v'$ that is a copy of some vertex $v$, we put $b(v')=1$. 
\end{definition}

We define the notion of a {\bf fragment}, that is to denote a possible fragment of an alternating cycle from $C \oplus C'$  
contained in a bad cycle.  Let $v_1 \neq v_2$ belong to bad cycle $c_i$  from $C$. Then the fragment connected with $v_1, v_2$  is any alternating path
$(v_1,v_3,v_4,...,v_k,v_2)$ whose all vertices belong to $c_i$ and such that it begins and ends with an edge in $C$.  Thus, if $c_i$ is a triangle and $v_1, v_2$ are its two different vertices, then the fragment corresponding to them is the edge $(v_1, v_2)$.

\begin{figure} 
\centering{\includegraphics[scale=0.6]{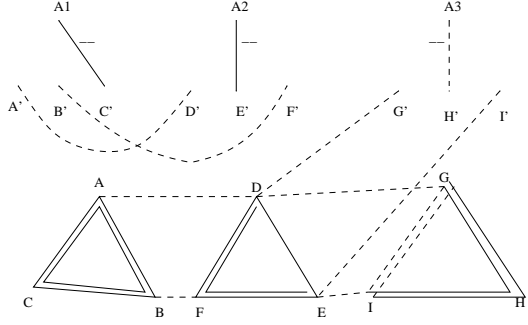}}
\caption{\scriptsize The weight of the edges drawn with a solid line is $5$, with a dashed line $3$ and a minus near the edge shows that the edge has negative weight ($-5$ or $-3$). The weight of the remaining edges of $G$ is $0$. Cycle cover $C$ of maximum weight
consists of cycles $ABC, DEF, GHI$, a good cycle cover $C'$ consists of cycle $ACBFEIHGD$ and a $b$ matching $B$ of maximum weight in $G'$ consists of cycles $ABC, GHI$ and path $G'DFEI'$ and edges $A1C',A2E', A3H', A'D', B'F'$.
$C \oplus C'$ consists of two alternating cycles $(AD,DF,FB,BA)$ and $(DG,GI,IE,ED)$. Their alternating weight equals to correspondingly $-4$ and $-2$. $C \oplus B$ consists of edges $A1C',A'D',B'F',A2E'$ and $A3H',G'D,DE,EI'$.
Let us notice that the alternating weights of these sets of edges are also $-4$ and $-2$ and that the weight of $C'$ and $B$ are the same.} \label{gprim2}

\end{figure}

A $b$-matching of $G'$ is defined in such a way that for each good cycle cover $C_1$ of $G$, we are able to find
a $b$-matching $B$ of $G'$ that corresponds to it in the sense that alternating cycles $C'\oplus C$ are virtually the same as alternating cycles $B \oplus C$. (These are not quite alternating cycles.)  
Informally speaking, parts of alternating cycles from $C \oplus C'$  
contained in bad cycles correspond in $G'$ to the edges contained in the gadgets and the remaining parts of the alternating cycles are in a way impressed in the graph $G$.

A $b$-matching of $G'$ is  such that for each bad cycle $c_i$ there are exactly two vertices, say $v_1,v_2$, such that $v_1',v_2'$  will be matched with the edges  not contained in the gadget $U_i$ (these edges will be of the form $(v_1',v_3), (v_2',v_4)$) and the weight of the edges contained in $U_i$ corresponds to the alternating
weight of the fragment connected with $v_1$ and $v_2$.

We will say that a $b$-matching $B$ of $G'$ {\bf lies by an error $\epsilon \geq 0$} on a bad cycle $c_i$ if for  vertices $v_1'$, $v_2'$  (such that $v_1,v_2$ belong to $c_i$) matched with edges not contained in a gadget $U_i$, the weight $w_i$ of the edges of $B$ contained in $U_i$ satisfies the following inequality: 
$w'(f_i) \geq w_i \geq w'(f_i)-\epsilon w(c_i),$
where $f_{i}$ denotes some fragment connected with $v_1, v_2$ and $w'(f_i)$ its alternating weight.

We will prove
\begin{lemma} \label{error}
Every $b$-matching $B$ of $G'$ lies on a bad triangle by an error $0$ and on a bad square by an error at most $\frac{1}{18}$. 
\end{lemma}
The proof is given in Appendix.

\begin{figure} 
\centering{\includegraphics[scale=0.6]{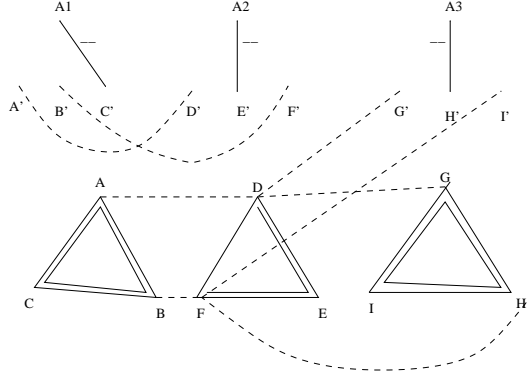}}
\caption{\scriptsize The weight of the edges are represented in the same way as in Figure \ref{gprim2}. A cycle cover $C$ of maximum weight consists of cycles $ABC, DEF, GHI$. A $b$-matching $B$ of maximum weight consists
of cycles $ABC,GHI$, path $G'DEFH'$ and edges $A3I',A2E',A1C',A'D',B'F'$. $S_B=\{DA,AB,BF,FD,FD,DG,GH,HF\}$(there is a mistake in the figure).} \label{bmatch}
\end{figure}

Clearly $B$ in $G'$ is {\it not} a good cycle cover of $G$ (it is not even a cycle cover of $G$). Let us however point the analogies between $B$ and a good cycle cover of $G$.
Let us define for $B$ a {\bf quasi-alternating} multiset $S_B$. $S_B$ will contain:  (1) for each edge $e=(v_1,v_2)$ \ $|(Z_e \cap B)\setminus C|$ number of copies of $e$, where  $Z_e=\{(v_1,v_2), (v_1',v_2'), (v_1,v_2'), (v_1',v_2)\}$, (2) the set of edges $C \setminus B$, (3) for each gadget $U_i$ it contains a fragment connected with $v_1,v_2$ iff $v'_1,v'_2$ are matched in $B$ with vertices from the original graph $G$. For example in Figure \ref{gprim2} $S_B=C\oplus C'$. Another example is given in Figure \ref{bmatch}.  

 The alternating weight of a multiset is defined in an analogous way so that the weight of the edge not in $C$ is counted the number 
of times it occurs in the multiset and the weight of the edge in $C$ is subtracted the number of times it occurs in the multiset.
We have
\begin{fact} \label{quasi}
$$ w(B)=w(C)+w'(B\oplus C).$$
\end{fact}
Next, we are going to bind good cycle covers with $b$-matchings in $G'$.

\begin{lemma} \label{bind}
If $C_1$ is a good cycle cover, then there exists such a $b$-matching $B$ in $G'$ that
$w'(S_B)\geq w'(C_1\oplus C)$.
\end{lemma}

By Fact \ref{przek} for each cycle $c$ of length $3$ or $4$ there exists an alternating cycle $K_c$ in $C \oplus C_1$ such that  there are two different vertices $v_1, v_2$ on $c$ such that the part of an alternating cycle $K_c$ between
vertices $v_1, v_2$ is a fragment (i.e. this part is on vertices solely from $c$). 

  For some cycles there are more 
such alternating cycles or there is more than one place of this kind on such an alternating cycle. Nevertheless for each cycle $c$ we choose
one such $K_c$ and one subpath $P_c$ on it. Next we build a $b$-matching $B$. Originally let all the edges
from $C$ belong to $B$. Each subpath $P_c$ is encoded by the corresponding gadget
and for the remaining edges of $K_c$ we do as follows. If $e \in E \setminus C$, then we add $e$ to $B$ 
(or more precisely sometimes a corresponding edge between the copies of vertices) and if $e \in C$, then we remove it from $B$.  
For example in Figure \ref{gprim2} for cycles $ABC$ and $DEF$ we chose an alternating cycle $(AD,DF,FB,BA)$ and for cycle $GHI$ the other alternating cycle.
We could also choose for cycle $ABC$ the same alternating cycle $(AD,DF,FB,BA)$, but for $DEF$ and $GHI$ the other one. Then $b$-matching $B$ would consist of cycles $ABC,GHI$, path $A'FEDC'$ and edges $A1C',A2E',A3H', D'G',F'I'$.

By Lemma \ref{error} we have that $w'(B\oplus C) \geq w'(S_B)-\frac{1}{18}w(C)$ and therefore we get
\begin{corollary}
$w(B) \geq w(C') - \frac{1}{18}w(C)$. Recall that $C'$ denotes a good cycle cover of maximum weight.
\end{corollary}

From $B \cup C$ we obtain a $4$-regular graph $H$. We do it in the following way. At the beginning $H$ consists of two copies of the cycle cover $C$ (at this moment $H$ is $4$-regular). Next we compute $S_B$ and apply it to $H$, that is we put $H:=H \oplus (B \oplus S_B)$.

For example in Figure \ref{bmatch} we would get $H=\{AC,AC,CB,CB,AB,AD,BF,FE,\\
FE,ED,ED,DG,FH,GI,GI,IH,IH,GH\}$ and in Figure \ref{gprim2} $H=\{AC,AC,CB,CB,\\
AB,AD,BF,DF,DE,FE,FE,EI,DG,GI,IH,IH,HG,HG\}$.

\section{Extracting a heavy tour} \label{heavy}

To {\bf $2$-path-color} graph $G$ will mean to color its edges into two colors (each edge is colored
into one color) so that the edges of the same color form a collection of node-disjoint paths.  

To {\bf $2$-cycle-color} the graph will mean to color its edges into two colors so that
the edges of each color form a collection of node-disjoint cycles.
Since the graph can contain double edges, some of these cycles can be of length $2$.
To {\bf well $2$-cycle-color} the graph  will mean to $2$-cycle-color  it so that each monochromatic
cycle has length at least $5$.

Since $H$ is $4$-regular, we can $2$-cycle-color it. If we could well $2$-cycle-color it, then we would 
put the edge of minimal weight from each monochroamtic in $E'$, then $E'$ would have weight at most $1/5 w(H)$ 
and graph $H'=(V,E_H\setminus E')$ would be $2$-path-colored.
As one can easily check, however, there exist graphs that cannot be well $2$-cycle-colored.

We can however restrict ourselves to considering graphs that (almost) do not contain triangles as we prove
Lemma \ref{triangle}, which is the corollary of two lemmas from Section \ref{tr}. 
\begin{lemma} \label{triangle}
In Theorem \ref{koloro} we can restrict ourselves to graphs $H$ such that if a triangle $T$ is a subgraph of $H$,
then either (1) $T$ contains two double edges or (2) $T$ consists of single edges and each vertex of $T$ is adjacent
to a double edge.
\end{lemma}
(If we can eliminate a triangle from a connected component having $5$ vertices using lemmas from Section \ref{tr},
then we do not do that but deal with such a component separately.)
It would be nice to be able to restrict ourselves also to graphs that do not contain cycles of length $2$
or $4$.   However, we have not been able to find an analogous way to that from lemmas in Section \ref{tr}.
Instead we will well $2$-almost-cycle-color the graph, which we define as follows.
To {\em $2$-almost-cycle-color} the graph means to color
the subset of its edges into two colors, so that the edges of each color form a collection of node-disjoint paths
and cycles and the set of uncolored (called {\em blank}) edges is node-disjoint. (The set of blank edges can be empty.)
To {\em well $2$-almost-cycle-color} the graph means to $2$-almost-cycle-color 
it so that each monochromatic cycle has length at least $5$.   

In Section \ref{dis}, we give the algorithm for well $2$-almost-cycle-coloring the graph. The key part of the algorithm 
is played by {\em disabling} cycles of length correspondingly $2$, $3$ and $4$, which consists in such a colouring 
of a certain subset of the edges that whatever happens to the rest of the edges no monochromatic cycle of lengh $2$, $3$ or $4$ will  arise. 

Once graph $H$ gets well $2$-almost-cycle-colored, we would like to find such a subset $E'$ that $w(E')\leq 1/5 w(H)$
and such that after the removal of $E'$ from $H$, $H'=(V,E \setminus E')$ is $2$-path-colored that is the edges that got colored
in $2$-almost-cycle-coloring keep their color and blank edges are colored into an appropriate color.

In Section \ref{par} we will describe five phases of dealing with a well $2$-almost-cycle-colored graph $H$:
two red ones, two blue ones and one blank one.
With the phases we will attach five disjoint subsets of edges $R_1,R_2,B_1,B_2,Blank$ such that
in the $i$-th ($i=1,2$) red phase we will obtain a graph $P_{R_i}=(V_t,E_t \setminus R_i)$, which after
coloring the remaining blank edges red, will be $2$-path-colored, analogously for the blue phases.
In the blank subphase we will obtain a graph $P_{Bl}=(V_t,E_t \setminus Blank)$, which after coloring
the remaining blank edges into an appropriate color will also be $2$-path-colored. 
Thus each blank edge acts twice (i.e. in two phases) as a red edge, twice as a blue edge and once it is removed.

\section{Eliminating triangles} \label{tr}
\begin{definition}
Suppose we have a graph $J=(V_J,E_J)$ as in Theorem \ref{koloro} (i.e. loopless, $4$-regular,
having at most $2$ edges between a pair of vertices and such that each of its connected components has at least $5$ vertices) and its subgraph $S$.
We say that we can {\em eliminate} $S$ from $J$ iff there exists graph $K=(V_K,E_K)$ that does not contain $S$, has at least one vertex less than $J$ and such that the solution from Theorem \ref{koloro} for $K$ can be transformed into a solution for $J$, which means that if we have a set $E_K' \subset E_K$ such that $w(E_K')\leq 1/5 w(K)$ 
a $2$-path-coloring of graph $K'=(V_K, E_K \setminus E_K')$, then we can find a set $E_J'$ such that $w(E_J') \leq 1/5 w(J)$ and such that graph $J'=(V_J,E_J \setminus E_J')$ can be $2$-path-colored.
\end{definition}

First, we will eliminate triangles that contain exactly one double edge.
\begin{lemma} \label{podw}
If a triangle $T$ has exactly one double edge, then we can eliminate $T$. 
\end{lemma}

The proof is given in Appendix.

Next, we will eliminate triangles that do not contain any double edges.

\begin{lemma}
If graph $J$ does not contain triangles having exactly one double edge, but contains a triangle $T$, whose all edges are single and such that at least one vertex of $T$ is not adjacent to a double edge,
then we can eliminate $T$ from $J$.
\end{lemma}

\begin{figure}[h]
\includegraphics{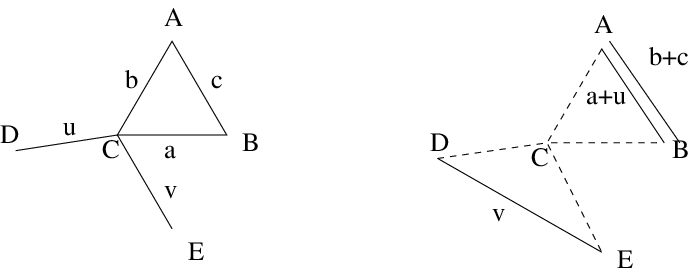}
\end{figure} \label{trojkat}

Suppose that  a triangle $T$ is on vertices $A,B,C$ and let $a,b,c$ denote the weights of the appropriate edges.  
First assume that $C$ is not adjacent to a double edge and $c\geq \min\{a,b\}$.
Thus $C$ is also connected by two edges with two other vertices $D,E$ and $u,v$ are the weights of these edges.
Without loss of generality, suppose that $u \leq v$.
We build $K$ as follows. At the beginning, it is the same as $J$.
Next, in $K$ we remove $C$ and all four edges adjacent to it.
Now $A,B,D,E$ are of degree $3$. We connect $D$ and $E$ with an additional edge of weight $u$. (Since $J$ does not
contain triangles with exactly one double edge, in $K$ $D$ and $E$ will be connected with at most two edges.)  We connect $A$ and $B$ with an additional edge of weight $a+v$ and we change the weight of the edge
that had weight $c$ to $c+b$. Note that $w(K)=w(J)$.

Assume we have found the set of edges $E_K'$ and have $2$-path-colored $K'$.
Except for edges $b+c,a+v,u$, $E_J'$ will contain the same edges as $E_K'$ and the $2$-path-coloring of $J'$
will differ from that of $K'$ only on these edges.

If from $K$ none of the edges $c+b,a+v,u$ was removed then, in $K'$ vertices $A$ and $B$ are connected with a double edge and suppose that the edge $DE$
is blue, then in $J'$ we color $CD$ and $CE$ into blue and $AC,CB$ into red.
If the edge $u$ was removed and edges $b+c,a+v$ not, then $E$ is in $K'$ adjacent to at most three edges and suppose that
two of them are blue. Then, we color $EC$ into red and $AC,CB$ into blue and $AB$ is left as a red edge.
If the edges $a+v,b+c$ were removed and the edge $u$ not and is blue, then in $J'$ we color $CD,CE$ into blue.
If $a+v \leq b+c$ and the edge $a+v$ was removed but the edges $u,b+c$ not, then
\begin{itemize}
\item if $b+c$ is red and $u$ blue, then if  $A$ has only one red edge incident on it ($b+c$), color $AC$ into red,
$DC$ into $blue$, else if $A$ has two red edges incident on it but
is not connected with $D$ via a blue path that does not contain $u$, then color $AC,CD$ into blue (notice that we do not create a blue cycle), else if $A$ is connected with $D$ via a blue path that does not contain $u$, color $AC,EC$ into blue (notice that $E$ is not connected with $A$ via a blue path not containing $u$).
\item if $b+c,u$ are blue, then if $A$ has at most one red edge incident on it, color $AC$ into red and $DC$ into blue,
else if $A$ has two red edges incident on it, color $AC$ into blue and either $EC$ or $DC$ into blue so as not to create
a cycle (since at most one of the vertices $D,E$ is connected via a blue path not containing $u$ with $B$, it is always possible).
\end{itemize}

If $a+v>b+c$ and the edge $b+c$ was removed but the edges $a+v,u$ not, then if $u$ is blue, color $CD,CE$ into blue.
Since $c \geq a$, it is all right. 

If $c < \min\{a,b\}$ and both $A$ and $B$ are adjacent to double edges, then everything goes as above. The only trouble could arise if $a+v>b+c$ and the edge $b+c$ was removed but the edges $a+v,u$ not and $a+v,u$ are blue.
Now we cannot only color $CD,CE$ into blue and not take any of the edges $AC,BC$. However, since $B$ is adjacent to a double edge, it has at most one red edge incident on it, so we can additionally color $BC$ into red.

Let us notice that $w(E_J') \leq w(E_K')$.

The rest of this section is given in Appendix.

{\bf \Large Appendix}

\section{Algorithm for well $2$-almost-cycle-coloring} \label{dis}
By Lemma \ref{tr} we can assume that, if graph $H$ contains a triangle $T$, then either it has two double eges
or it consists only of single edges and each vertex of it is adjacent to a double edge.
We will give the algorithm for well $2$-almost-cycle-coloring $H$. We will use colors: blue and red.
In the algorithm once the edge gets colored, it will not change its color and we will preserve the following invariant.
\begin{invariant}
If at some step of the algorithm exactly two of the edges incident on vertex $v$ are colored, then they have different
colors.
\end{invariant}

We remind that disabling cycles of length less than $5$ consists in such a coloring of the subset of the edges of $H$, that however the rest of blank edges are coloured, the graph will
not contain a monochromatic cycle of length less than $5$.
We begin from cycles of length $2$. Disabling such cycles is very easy, we consider each double edge
and colour it into two different colours: red and blue. As the graph does not contain connected components having
less than $5$ vertices, no monochromatic cycle of length less than $5$ will arise.
Next we disable caps. {\em A cap} is a triangle that has exactly two double edges or a square that has exactly three double edges.
Let $(v_1,v_2)$ denote the only non-double edge of a given cap $C$. Then the non-double edges incident at $v_1$ and $v_2$ different from $(v_1,v_2)$ are called the {\em ribbons} of cap $C$. A ribbon may belong to two different caps.

We can eliminate some caps from the graph in a way similar to that in which we eliminated most kinds of triangles in the previous section.
\begin{lemma}
If a cap $C$ with two ribbons $r_1,r_2$ is such that $r_1,r_2$ do not share a vertex and are not connected by a double edge, then we can eliminate $C$.
\end{lemma}
If a cap $C$ with two ribbons $r_1,r_2$ is such that $r_1,r_2$ share a vertex or are connected by a double edge, then
we disable it by coloring $r_1,r_2$ into different colors.

We will say that a square or triangle is {\em active} if some of its edges (possibly all)
are blank and there exists such a well $2$-almost-cycle-coloring of all blank edges in the graph that this square
or triangle is monochromatic. 

We say that edge $e$ is {\em active} if it is colored and included in some active square.
In disabling squares we will maintain the following property of active edges.
\begin{invariant}
If edge $e=(v_1,v_2)$, say red, is active, then $v_1$ has either two or four coloured edges incident to it
and the same for $v_2$ and either $e$ is double or neither $v_1$ nor $v_2$ has an active blue edge incident at
it. 
\end{invariant}

 Consider the following algorithm.

{\bf while} there are active squares do \\
{\bf if} there is an active square $s$ with two active edges, then color the two blank edges of $s$ into different colors. 
  
 {\bf else if} there is an active square $s=(v_1,v_2,v_3,v_4)$ with one active edge $e=(v_1,v_2)$ (double or say, to fix the attention red), then check if there is another active square $s'=(v_1,v_2,v_3',v_4')$ that contains $e$ but no other edge of $s$. Notice that since $H$ does not contain triangles, $s'$  cannot contain $v_3$ or $v_4$.  
 Color $(v_2,v_3)$ and $(v_1,v_4)$ into red and $(v_3,v_4)$ into blue. If $s'$ exists and $(v_1,v_2)$
is double color $(v_2,v_3')$ and $(v_1,v_4')$ into blue and $(v_3',v_4')$ into red (otherwise do nothing,
as it is not needed).

{\bf else if} there is an active square $s$ with all blank edges, then
  color the edges of $s$ alernately into blue and red.
Next if the red edge $e_2$ belongs to an active square $s_1$, color the edges of $s_1$ adjacent to $e_2$ into red
and the remaining one into blue. Next if the blue edge of $s$:$e_1$ or $e_3$ belongs to an ative square, color it analogously: two edges into blue and one into red. Notice that if $e_1$ or $e_3$ belongs to an active square then it is not adjacent on the blue edge of $s_1$, because the graph does not contain triangles. Next, do the same with the remaining colored edges of $s$, if they belong
to an active square.   

\section{Partition} \label{par}

In this phase  we will give the algorithm that finds five disjoint subsets of edges $R_1,R_2,B_1,B_2,Blank$ 
corresponding to the five subphases: two red ones, two blue ones and one blank one such that after removing the edges
from each one of these and coloring the blank edges, depending on the subphase: red (in the red subphases) or blue (in the blue subphases), the graph will contain only blue or red paths, that is will be $2$-path-colored.  

After disabling cycles of length less than $5$, we arbitrarily color the rest of the edges, as a result in the graph
no two adjacent blank edges will be left.  
Thus if we have a blank edge $e$ between vertices $v_1$ and $v_2$, then the remaining three edges of $v_1$ are coloured: two 
into blue and one into red and the remaining three edges of $v2$ are coloured: two into red and one into blue or vice versa.
We will say that the blue edges of $v_1$ and the red edges of $v_2$ are the blue or red {\em heads} of the edge $e$.  
We will also say that a red edge of $v_1$ is the red tail of $e$ and a blue edge of $v_2$ is a blue tail of $e$.
Let us notice that if we would like to colour a given blank edge blue or red, then we have to remove one 
blue or correspondingly red head.  
From the point of view of a given blank edge $e$ the situation presents itself as follows.
In the red subphases it is coloured red and one of its heads is removed in one red subphase and the other head is removed in the second subphase, thus one of its heads must belong to $R_1$ and the other one to $R_2$ and 
analogously in the blue subphases. In the blank subphase $e$ is simply removed.
Therefore we can see that a blank edge and its four heads fall into five different sets $R_1,R_2,B_1,B_2, Blank$.
We will call all the blank edges and their heads {\em charged (edges)}.
In the red and blue subphases we must be careful not to create cycles that consist solely of charged edges.
We are not allowed to create such cycles, because we cannot afford to remove any edge from this cycle,
as all of them must belong to the sets attached to other subphases (for example, $R_2, Blank$ if we are now in the first
red subphase).
\begin{lemma}
If the (blue or red) cycle $c$ consists only of charged edges, then for every blank edge belonging to $c$
we have that its head (the one that belongs to $c$) is a tail of another blank edge belonging to $c$.  
\end{lemma}

Suppose that $c$ contains $k$ (originally) blank edges. First let us notice that $k>1$.
Since $c$ consists only of charged edges, all the edges between two consecutive blank edges on $c$ must be charged.
There are exactly $k$ disjoint nonemptysubsets of edges connecting the $k$ blank edges on the cycle. Let us fix one direction
of movement along $c$, say clockwise. Then, each blank edge is either followed or preceded by its head.
Let us observe, that if $c$ consists only of charged edges we must have that either each blank edge is followed
by its head or each blank edge is preceded by its head, because otherwise one set of edges connecting certain two
blank edges would not contain a charged edge. Since all the edges must be charged, the sets connecting the blank edges
must contain one edge each.

\subsection{Preprocessing}

In this phase we want to eliminate as many as possible of the following two situations:
\begin{enumerate}
\item  a coloured edge is charged by two different blank edges
\item  a blank edge is incident to a blue or red cycle.
\end{enumerate}

The algorithm for preprocessing and the exact description of what is eleiminated in the graph is given in Appendix.

\subsection{Algorithm of Partition}

The main problem consists in partitioning the red and blue heads of each blank edge into $R_1, R_2, B_1, B_2$
so that no cycle consisting only of charged edges arises. If we succeed with this, we can show
that we can add some non-charged edges to $R_1,R_2,B_1,B_2$ depending on the phase so that their removal completely decycles the graph.

We will say that a blank edge $e'$ is a {\em descendant} of a blank edge $e$ if the tail of $e$ is a head of $e'$
or if the tail of $e$ is a head of a blank edge $e''$ such that $e'$ is a descendant of $e''$.
%We will also assume that a blank edge is it own descendant.
%A blank edge $e$ is an {\em ancestor} of $e'$ if $e'$ is a descendant of $e$.

We say that a blank edge $e$ is {\em twinny} if one of its heads is also a head of another blank edge $e'$. (%and the other heads of $e$ and $e'$ (both) are the tails of some other blank edges $e'',e'''$). 
$e'$ is called {\em a twin} of $e$. 
The common head of two twinny edges is called the {\em comhead}.

Let us now describe the algorithm of choosing which head of each blank edge falls into $R_1$(correspondingly $B_1$)
and which one into $R_2$ (correspondingly $B_2$).
Let us notice that if we decide which head falls into $R_1$ ($B_1$), we automatically know that the other one will have to fall into $R_2$ ($B_2$). 
Since the algorithm is the same for red and blue edges, we will describe it for red edges.
We assume that at the beginning  all blank edges are uncolored.
At each step of the algorithm we choose one uncolored blank edge $e$ and color it into red (now $e$ is said to be colored) and decide which of its heads falls into $R_1$. If $e$ is twinny and the chosen head is the comhead, then we also color $e'$, which is a twin of $e$ into red.
We say that a cycle $c$ is {\em forbidden in the $i$th red phase} if it consists only of charged edges that do not fall into $R_i$. 

We say that a colored blank edge $e$ is {\em safe} if it has no chance to be a part of a forbidden cycle in the $2$nd red 
phase, no matter what happens to the rest of (yet uncolored) blank edges (i.e. no matter which heads of the rest
of uncolored  blank edges fall into $R_1$ and which into $R_2$).

\begin{fact}
A colored blank edge is safe if its tail falls into $R_2$ or is not a head of any blank edge or if its head that falls into $R_1$ is not a tail of any blank edge or if some descendant of $e$ is safe.
\end{fact}

We say that a colorod blank edge $e$ is {\em unprocessed} if its tail is a head of a yet uncolored blank edge.

We say that a colored blank edge $e$ is {\em exposed} if it is unprocessed and 
and there is a risk of creating a forbidden cycle in the $1$st red phase that includes $e$.

\begin{fact}
A colored blank edge $e$ is exposed if there exists  a colored blank edge
$e'$ such that $e$ is a descendant of $e'$ and all desendants of $e'$ (  that are not also descendants of $e$) are colored and $e'$'s head that does not fall into $R_1$ is a tail of some uncolored blank edge.
\end{fact} 

The blank edge that is not colored in the $2$nd red (blue) phase is called {\bf postponed (in the red (appropriately blue) phase)}.  
They will be colored in the Blank phase.

{\tiny

{\bf while} there are uncolored blank edges {\bf do}\\

{\bf If} there is an unprocessed unsafe edge $e$, then take an ucolored edge $e'$, whose head is a tail of $e$.\\
{\bf Otherwise if} there is an exposed edge $e$, then take an ucolored edge $e'$, whose head is a tail of $e$.\\
{\bf Otherwise} take an arbitary uncolored blank edge $e'$. \\

Let the heads of $e'$  be called $h_1$ and $h_2$ and let $h_1$ denote that head, which is also a tail of $e$.\\

{\bf if $e'$ is not twinny}, then check if coloring $e'$ and putting $h_2$ into $R_1$ would create a forbidden cycle in the $1$st phase. \\
If the answer is ''no'', then put $h_2$ into $R_1$ and $h_1$ into $R_2$ and color $e'$. \\ 
If the answer is ''yes'', then put $h_1$ into $R_1$ and $h_2$ into $R_2$ and color $e'$.  \\

{\bf if $e'$ is twinny}, then $h_2$ is a comhead of $e'$ and its twin $e''$. Let the other head of $e''$ be called $h_3$.
Check if coloring $e', e''$ and putting the comhead $h_2$ into $R_1$ would create a forbidden cycle in the $1$st phase. ($e'$ and $e''$
can create two different forbidden cycles or $e'$ can create a forbidden cycle but $e''$ not or vice versa.)
If the answer is ''no'', then put $h_2$ into $R_1$ and $h_1$ into $R_2$ and color $e',e''$. \\
If the anser is ''yes'', then put $h_1$ and $h_3$ into $R_1$ and color $e',e''$. \\ 
If putting $h_2$ into $R_2$ and coloring $e',e''$ would create a forbidden cycle in the $2$nd red phase, then call $e',e''$ postponed in the red phase,
otherwise put $h_2$ into $R_2$.

}

\begin{lemma}
We do not create a forbidden cycle either in the $1$st or $2$nd red or blue phase and we do not create any cycles in the $Blank$ phase. No blank edge is postponed both in the red and blue phase.
\end{lemma}

At each step of the algorithm the following holds: 
{\bf Claim}
{\it No forbidden cycle has been created in the $1$st or $2$nd red or blue phase and no cycle has been created in the $Blank$ phase and there is at most one unsafe edge. The unsafe edge,
if exists at a given step, is unprocessed.}

Before the algorithm starts, Claim is true. In the first step we take an arbitrary blank edge $e'$. If $e'$ is twinny, then we put the comhead of $e'$ and the twin $e''$ into $R_1$, thus afterwards both $e'$ and $e''$ are safe
(as the colored path containg $e',e''$ contains a comhead in the $2nd$ red phase).  It is also not possible to create a forbidden cycle in the $1$st phase containing $e'$ or $e''$. However $e',e''$ may be exposed.
If $e'$ is not twinny, then we put one of its heads into $R_1$ and the other one into $R_2$. We do not create a forbidden cycle either in the $1$st or $2$nd phase. $e'$ may be unsafe, but if it is unsafe, then it is also
unprocessed.

Suppose that till the $k$th step Claim is true. We perform another step.
We deal with an uncolored blank edge $e'$. We have a few cases.\\
{\it Case 1} If there is an unsafe edge $e$, then one of $e'$'s heads is a tail of $e$. 
{\it Case 1a} $e'$ is not twinny. Then if we put $h_2$ into $R_1$, then we do not create a forbidden cycle in the $1$st phase and afterwards $e$ is safe. $e'$ may be unsafe but we do not create a forbidden cycle in the $2nd$ phase as now $e'$ is the only possibly unsafe edge (and the forbidden cycle in the $2$nd phase must contain at least two unsafe edges.)  If we put $h_1$ into $R_1$, then it means that putting $h_2$ into $R_1$ and coloring $e'$ red would create a forbidden cycle in the $1$st phase. Suppose that $e'=(vh,vt)$  and $vh$ is the vertex that has the heads of $e'$ incident on it. It means that $vt$ is the end of the red path ending on $h_1$ in the $1st$ red phase. Thus if we put $h_1$ into $R_1$, then we do not create a cycle containg $e'$ in the $1$st red phase. As for the $2$nd red phase it means that the tail of $e'$ is not present there (because it is present in the $1$st phase), therefore $e'$ will be the end of the red path. \\

{\it Case 1b} $e'$ is twinny and $e''$ is its twin.  
If we put the comhead $h_2$ into $R_1$, then we do not create a forbidden cycle in the $1$st red phase and afterwards $e,e',e''$ are safe - $e',e''$ are safe because the comhead $h_2$ is present in the $2$nd red phase.
If we put $h_1,h_3$ into $R_1$ and $h_2$ into $R_2$,  then it means that doing otherwise (putting $h_2$ into $R_1$) created a forbidden cycle in the $1$st phase containing $e'$ or $e''$. If $e'$ would have been in the forbidden cycle in the $1$st red phase, then now it becomes safe (for the same reasons as that in case 1a) and thus at most one edge remains unsafe.
If we put $h_1,h_3$ into $R_1$ and postpone $e',e''$, then it means that putting $h_2$ into $R_1$ creates a forbidden cycle in the $1$st phase and putting $h_2$ into $R_2$ creates a forbidden cycle in the $2$nd red phase, which
implies that $e'$ would belong to a forbidden cycle in the $2$nd  phase and $e''$ to a forbidden cycle in the $1$st phase, which in turn implies that in the Blank phase we do  not create a cycle containing $e'$ or $e''$.

the remaining cases are similar and easier.

\begin{fact} \label{releg}
If edge $e_r$ is a red head of the postponed blank edge $e$, then in the $2$nd red phase $e_r$ is not a part of any red cycle. The same is of course true for the $2$nd blue phase.
\end{fact}

Consider all red cycles that are created in the two red phases (they are not forbidden), so each one of them
contains at least one uncharged edge. Let us build a bipartite graph $H_r=(C \cup E_{ur}, E')$ that has red cycles $C$ on one side and uncharged
red edges $E_{ur}$ on the other side and we connect a red cycle $c$ with the uncharged edge $e_r$ by an edge iff $e_r$
belongs to $c$. We wish to find a matching in $H_r$ in which every cycle $c$ is matched.

\begin{lemma}
In $H_r$ (and $H_b$ alike) there exists a matching $M_r$ of the cardinality $|C|$
\end{lemma}

If some edge $e_r$ is adjacent to some red head $h_e$ of a blank edge $e$, then its degree in $H_r$ is equal to $1$,
because if $e_r$ is a part of a cycle $c$ in the $i$th phase, then $e_r$ is not a part of a cycle in the $3-i$th phase,
as if $e$ is colored in the $3-i$th red phase, then $h_e$ falls into $R_{3-i}$ beacuse it was present in the $i$th phase
or $e$ is not colored in the $3-i$th phase (because it will be colored in the $Blank$ phase), but then by Fact \ref{releg} $h_e$ is not a part of any red cycle in the phase in which it remains blank.

Each edge $e_r$ has degree $2$ at most in $H_r$ as it can belong to at most one red cycle in each of the two phases.

Suppose that a matching $M_r$ of the cardinality $|C|$ does not exist in $H_r$. Then there exists such a subset of cycles $C'$ that a set $N(C')$ of the uncharged edges belonging to cycles from $C'$ has a smaller cardinality than $|C'|$, i.e. $|N(C')| < |C'|$. Without loss of generality, assume that $C'$ is a minimal (under inclusion) such set.
It means that in $C'$ there are exactly two cycles $c_1,c_2$  that  have degree $1$ in $H_r$ and if there are any more cycles in $C'$, then they have degree $2$ in $H_r$ and in $N(C')$ all uncharged edges have degree $2$ in $H_r$.
Let us notice that if some uncharged edge $e_r$  is adjacent to another uncharged edge $e_r'$, then if $e_r$ belongs to
a cycle $c$, then so does $e_r'$ and thus $e_r$ cannot belong to $N(C')$. Therefore if uncharged edge $e_r$ belongs to $N(C')$, then
it is not adjacent to any red head and it is not adjacent to any other uncharged edge, therefore it must be a common tail of two different blank edges. But then the existence of a set $C'$ implies, that in $G'$ exists a red cycle
consisting only of charged edges (it goes through the vertices belonging to blank edges incident to uncharged edges from $N(C')$ but those one that are not adjacent to the uncharged edges), which cannot happen.

{\bf Proof of Lemma \ref{error}}\\

We give the proof for a bad square $c_i$.
Suppose that $l_1+l_3 \leq l_2+l_4$. Since we have a cycle cover of maximum weight $p_1+p_2 \geq l_1+l_3$, which implies that $l_2+l_4 \geq l_5 +l_6$.
If within a gadget vertices that are matched with $a1_{c_i}$ and $a2_{c_i}$ are $v_3',v_4'$, then the weight of the edges
within a gadget is equal to $-l_1$, which is equal to exactly the alternating weight of the fragment connected with 
$v_1',v_2'$. The proof is analogous  for the fragment connected with vertices $v_3',v_4'$. If within a gadget vertices $v_1',v_3'$ are matched with $a1_{c_i},a2_{c_i}$, then the weight of the edges within the gadget is equal to $-p_1$, which
is equal to the alternating weight of the fragment connected with $v_2, v_4$.
If within the gadget vertices $v_1', v_4'$ are matched with $a1_{c_i},a2_{c_i}$, then the weight of the edges within the gadget
is either $-l_1/2-l_2/2-l_3/2+l_4/2$  or $-p_1 -w(a1_{c_i}v_3') -p_2 -w(a2_{c_i}v_2'$. We check that since $l_2+l_4 \geq l_5+l_6$, then the weight of the edges within the gadget will always be equal to $-l_1/2-l_2/2-l_3/2+l_4/2$.
However the alternating weight of the fragment connected with $v_2,v_3$ is $-l_2$, so the difference in the weights
is $-(l_2+l_4)/2+(l_1+l_3)/2$. The weight of each edge in a bad square is $< \frac{2}{9}w(c_i)$, where $w(c_i)$ is the weight of the square, therefore the maximal difference between the weight of the pairs of edges is $1/9 w(c_i)$,
which means that the matching lies on a square by an error at most $\frac{1}{18}$.

{\bf Proof of Lemma \ref{podw}} \\

\begin{figure} \label{double}
\centering{\includegraphics{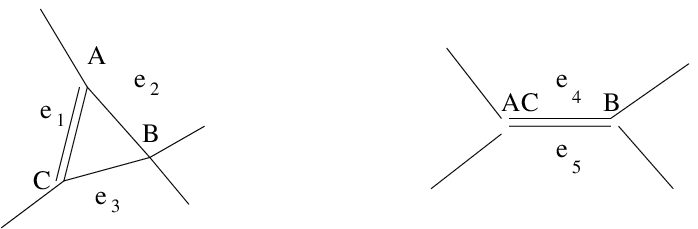}}
\end{figure}

Suppose that we have a graph $J=(V_J,E_J)$ that contains triangle $T$, whose  vertices  are $A,B,C$ and the double edge is between $A$ and $C$, then  we build graph $K$, which is the same as $J$, except for the fact that  in $K$ we shrink
vertices $A,C$ into one vertex $AC$ as shown in Figure \ref{double} (thus $K$ has one vertex less than $J$) and edges $e_1,e_1,e_2,e_3$ are replaced with edges $e_4,e_5$. In $K$ edge $e_4$ has weight equal to $w(e_1)+w(e_2)$
and edge $e_5$ has weight equal to $w(e_1)+w(e_3)$. Let us note that $w(K)=w(J)$.

Assume we have found the set of edges $E_K'$ and have $2$-path-colored $K'$.
Except for edges $e_4,e_5$, $E_J'$ will contain the same edges as $E_K'$ and the $2$-path-coloring of $J'$ will be different from $2$-path-coloring of $K'$ only on edges $e_1,e_1,e_2,e_3$. 
 If $e_4$ and $e_5$ are present in $K'$ (i.e. they do not fall into $E_K'$), then in $2$-path-coloring of $K'$ they have different colors and then in $J$ edges $e_1,e_1,e_2,e_3$ also do not fall into $E_J'$ and in $J'$ they are colored as  suggested by the coloring of $e_4,e_5$, i.e. if $e_4$ is colored blue, then $e_1$ (one copy of it) and $e_2$ are colored blue.  
If one or both of $e_4,e_5$ got removed in $K'$, then we remove (i.e.put into $E_J'$) $e_1,e_2$ if $e_4$ was removed  and we remove $e_1,e_3$ if $e_5$ was removed.  
Thus $w(E_J')=w(E_K')$.

{\bf Preprocessing}
To this end we will do some recolouring. At the same time we will take care not to create cycles of length less than $5$.

We will say that the edge $e$ (blank or coloured) has a blue or red roof if the colouring of this edge or the change
of its colour into a certain one is the cause of the arisement of a quadrilateral of this colour.   
 The part of this quadrilateral without edge $e$ is called its (blue or red) {\em roof}  The middle edge of the roof (consisting of $3$ edges) is called its {\em middle}.

\begin{lemma} \label{kwad}
If we cannot eliminate the situation that a coloured edge $e$ is charged by two different blank edges $e_1$ and $e_2$,
then at least one of the following conditions holds. Without loss of generality assume that $e$ is red.
\begin{enumerate}
\item  $e$ has a parallel edge,
\item  $e$ has a blue roof,
\item  $e_1$ has a red roof,
\item  $e_2$ has a red roof.
\end{enumerate} 
If none of these conditions holds we can eliminate the situation by changing the colour of $e$ into blue and colouring
edges $e_1$ and $e_2$ into red. We can also observe that if $e$ is a part of a red cycle, then neither $e_1$ nor $e_2$
can have a red roof.
\end{lemma} 

If we eliminate the situation as described above, then the colouring is of course correct. What can bother us is that cycles of length less than $5$ can arise. We do not have to worry about triangles. Also no cycle of length $2$ arises,
because $e$ does not have a parallel edge and edges $e_1$ and $e_2$ do not have parallel edges either because they were blank.
The edge $e$ cannot be a part of a blue quadrilateral because it does not have a blue roof. The edges $e_1$ and $e_2$ cannot both be a part of one red quadrilateral because it would mean that that there would have to be two additional red
edges as in Figure ??. However the first case cannot arise because $e$ is blue (after the elimination) and is not double
and the second case would imply that graph $G'$ contains triangles.

\begin{lemma} \label{cycle} 
If a blank edge $e$ is adjacent to a coloured cycle and at least one of its charged edges $e_c$ belonging to the cycle
is not double and is not charged by any other blank edge, then, suppose the cycle is blue, we can colour the blank edge $e$ 
blue and make the edge $e_c$ blank and no new cycles (of any length) will arise.  
\end{lemma}

No blue cycle can arise because $e$ becomes a part of a blue path, whose one end is adjacent to a blank (after the elimination) edge $e_c$.

Now we are ready to write the algorithm for the preprocessing phase.

{\bf Preprocessing}

		 {\bf while possible}\\
		   {\bf while possible} do the elimination from Lemma \ref{kwad} \\
			 {\bf if} some elimination from Lemma \ref{cycle} is possible, do it \\

\begin{lemma}			
The preprocessing phase runs in time $O(mn)$ (very roughly), where $m$ is the number of the edges and $n$ the number of vertices in $G'$.

\end{lemma}

The eliminations from Lemma \ref{kwad}  diminishes the number of blank edges by at least one.
The elimination from Lemma \ref{cycle} does not increase the number of blank edges and decreases the number of coloured cycles. Each elimination takes constant time. The number of coloured cycles is $O(n)$.

\end{document}